\documentclass{article}
\pdfoutput=1
\usepackage{proceed2e}
\usepackage{times}

\usepackage{amsmath, amssymb, times}
\usepackage{epsfig}
\usepackage{algorithmic}
\usepackage{algorithm}
\usepackage{amsthm}

\newcommand{\mathbs}{\boldsymbol}
\newtheorem{lemma}{Lemma}
\newtheorem{claim}{Claim}

\def\etal{{et~al.}}
\def\Lo{$L_1$~}

\def\Lp{$L_p$~}

\title{$L_1$ Projections with Box Constraints}

\author{
Mithun Das Gupta  \\
Epson Research and Development, Inc. \\
San Jose, CA.\\
\And
Sanjeev Kumar \\
Dept. Elec and Comp Engg. \\
University of California, San Diego \\
\And
Jing Xiao \\
Epson Research and Development, Inc. \\
San Jose, CA. \\
}

\begin{document}

\maketitle

\begin{abstract}
We study the \Lo minimization problem with additional box constraints. We motivate the problem with two different views of optimality considerations. We look into imposing such constraints in projected gradient techniques and propose a worst case linear time algorithm to perform such projections.
We demonstrate the merits and effectiveness
of our algorithms on synthetic as well as real experiments.
\end{abstract}

\section{Introduction}
In the domain of constrained optimization, it is well understood that $L_2$ norm penalty imposes smoothness constraint while the $L_1$ norm imposes sparsity~\cite{Ng04}. Lately, sparse representations have been shown to be extremely efficient in encoding specific kinds of data, mainly, obeying power decay law in some transform space e.g. DCT etc. Donoho~\cite{Donoho04} provided sufficient conditions for obtaining an optimal $L_1$-norm solution which is sparse. Recent
work on compressed sensing~\cite{Candes06,Tsaig06} further explores how $L_1$ constraints can be used for recovering a sparse signal sampled below the Nyquist rate.

$L_1$ regularized maximum likelihood can be cast as a constrained optimization problem. Although standard algorithms such as interior-point methods \cite{Tibshirani94,Koh07} offer powerful theoretical guarantees (e.g., polynomial-time complexity, ignoring the cost of evaluating the function), these methods typically require at each iteration the solution of a large, highly ill-conditioned linear system; which are potentially very difficult and expensive.

\begin{figure}[htbp!]
  \includegraphics[width=8cm,height=4cm]{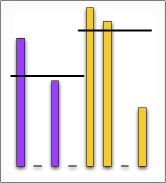}
  \includegraphics[width=8cm,height=4cm]{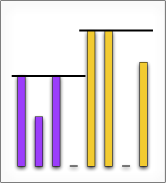}
  \caption{Left: \Lo projection and right: upper bounded \Lo projection for the same norm bound. Two colors represent different set of receivers with different upper bounds represented by black lines. \label{Fig:algo1_expl3434345}
}
\end{figure}

This paper explores the behavior of $L_1$ constraint optimization under the presence of bound constraints. Traditional $L_1$ constraint assumes an infinite upper bound on the magnitude of the predicates.
The idea of upper bounds can be explained very easily by a very simple example. Suppose we want to send a signal of length $n$ which is a combined signal originating from $k$ different sources. Suppose at the receiver side we have $k$ sets of receivers to decode the entire signal of length $n$. Based on the receiver set $i\in\{1,k\}$, the peak signal strength which the receivers can handle can be different. This kind of problem, 
is not handled by traditional $L_1$ projection. An illustration for the proposed problem, with 2 sets of receivers is shown in Fig.~\ref{Fig:algo1_expl3434345}. Assuming that transmitting 0's or upper bounds cost just 1 bit, the effective sparsity, assuming 8 bits per element, is (5*8+3*1)/64 for \Lo, and (2*8+6*1)/64 for upper bounded \Lo.
We motivate the inclusion of box constraints by looking at the problem from two different settings.

\textbf{Optimality Gap: } Let us consider an unconstrained problem
\begin{equation*}
    \min_\mathbf{x} \|\mathbf{x-v}\|_2^2 + \sum_i \lambda_i |{x}_i| + \mathbs{\zeta}^T(\mathbf{l-x}) + \mathbs{\gamma}^T(\mathbf{x-b})
\end{equation*}
where we have introduced the two bound constraints $[\mathbf{l,b}]$ into the cost function. This can be slightly modified to a function of 2 variables such that
\begin{eqnarray*}
  \min_{\mathbf{x,z}} && \mathbf{z}^T\mathbf{z} + \sum_i \lambda_i |{x}_i| + \mathbs{\zeta}^T(\mathbf{l-x}) + \mathbs{\gamma}^T(\mathbf{x-b}),\\
  s.t.&&  \mathbf{z} =  \mathbf{x-v}
\end{eqnarray*}
The Lagrangian for this problem can now be written as 
\begin{eqnarray*}
L(\mathbf{x,z},\mathbs{\beta})&=& \mathbf{z}^T\mathbf{z} + \sum_i \lambda_i |{x}_i| +\mathbs{\zeta}^T (\mathbf{l-x}) \\
&& + \mathbs{\gamma}^T(\mathbf{x-b}) + \mathbs{\beta}^T(\mathbf{x-v-z})
\end{eqnarray*}
The dual function is given by
\begin{eqnarray*}
&& \inf_{\mathbf{x,z}}L(\mathbf{x,z},\mathbs{\beta}) =  \inf_\mathbf{z}(\mathbf{z}^T\mathbf{z}-\mathbs{\beta}^T\mathbf{z}) + \\
 && \inf_\mathbf{x}(\sum_i \lambda_i |{x}_i| + (\mathbs{\beta-\zeta+\gamma})^T\mathbf{x}) \\
 &&-\mathbs{\beta}^T\mathbf{v}+\mathbs{\zeta}^T\mathbf{l}-\mathbs{\gamma}^T\mathbf{b} \\
   &=&
  \begin{cases}
   -\frac{\mathbs{\beta}^T\mathbs{\beta}}{4}-\mathbs{\beta}^T\mathbf{v}+\mathbs{\zeta}^T\mathbf{l}-\mathbs{\gamma}^T\mathbf{b} & \text{if } ({\beta-\zeta+\gamma})_i \le \lambda_i \\
   -\infty & \text{otherwise }
  \end{cases}
\end{eqnarray*}
The Lagrange dual can now be written as
\begin{eqnarray*}
  & \max & G(\mathbs{\beta}) = -\frac{\mathbs{\beta}^T\mathbs{\beta}}{4}-\mathbs{\beta}^T\mathbf{v}+\mathbs{\zeta}^T\mathbf{l}-\mathbs{\gamma}^T\mathbf{b} \\
  &s.t.& (\beta-\zeta+\gamma)_i \le \lambda_i
\end{eqnarray*}
A change of variables $\mathbs{\mu = (\beta-\zeta+\gamma)}$ leads to
\begin{eqnarray*}
  \max_{\mu_i \le \lambda_i} ~G(\mathbs{\mu}) &=& -\frac{\mathbs{\mu}^T\mathbs{\mu}}{4}-\mathbs{\mu}^T\mathbf{v} - \frac{\mathbs{\mu}^T(\mathbs{\zeta-\gamma})}{2} \\
  &-&\frac{(\mathbs{\zeta-\gamma})^T(\mathbs{\zeta-\gamma})}{4} \\
  &-&(\mathbs{\zeta-\gamma})^T\mathbf{v} +\mathbs{\zeta}^T\mathbf{l}-\mathbs{\gamma}^T\mathbf{b}
\end{eqnarray*}

Now the duality gap for the bounded problem can be written as
\begin{eqnarray*}
  \eta &=& \|\mathbf{x-v}\|_2^2 + \sum_i\lambda_i |x_i| + \\
  && \mathbs{\zeta}^T(\mathbf{l-x}) + \mathbs{\gamma}^T(\mathbf{x-b}) - G(\mathbs{\mu}) \\
   &=& \eta_{L1} - (\mathbs{\zeta - \gamma})^T(\mathbf{x-v})+ \\
   &&\frac{(2\mathbs{\mu+(\zeta-\gamma)})^T(\mathbs{\zeta-\gamma})}{4}
\end{eqnarray*}
where $\eta_{L1}$ is the duality gap for the \Lo problem without the bound constraints. For $x_i$ fixed at its upper bound $b_i$, $\mathbs{\gamma} > \mathbf{0}$ and $\mathbs{\zeta} = \mathbf{0}$. Under such a condition
\begin{equation*}
    \eta = \eta_{L1} - \mathbs{\gamma}^T(\mathbf{v-b})- \frac{(2\mathbs{\mu-\gamma})^T\mathbs{\gamma}}{4}
\end{equation*}
As long as $\mathbs{\gamma}^T(\mathbf{v-b}) + \frac{(2\mathbs{\mu-\gamma})^T\mathbs{\gamma}}{4} \ge 0$ the duality gap is reduced as compared to simple \Lo minimization. Since $\mathbs{\gamma}^T(\mathbf{v-x})$ is always positive (shown later in Sec.~\ref{SEC:UBSP}), a sufficient condition for reduction in duality gap is $\mathbs{\mu} \ge \frac{\mathbs{\gamma}}{2}$, where $\mathbs{\mu}$ is the dual feasible solution for the \Lo problem. The optimality of a particular solution is based on the duality gap and as such any decrement of the gap increases the optimality of the solution obtained. This shows that the optimality gap for the bound constrained \Lo problem can be made arbitrarily closer to zero compared to the similar unbounded \Lo problem.

\textbf{Degrees of freedom for upper bounded problem: }
We study the degrees of freedom of the upper bounded \Lo projection problem in the framework of Stein's unbiased risk estimation (SURE)~\cite{Stein81}. As shown by Zou \etal~\cite{Zou07} the number of non-zero coefficients is an unbiased estimate for the degrees of freedom (DF) of the optimization scheme.
The idea of upper bounds can be explained very easily by a very simple example. Suppose we want to send a signal of length $n$ which is a combined signal originating from $k$ different sources. Suppose at the receiver side we have $k$ sets of receivers to decode the entire signal of length $n$. Based on the receiver set $i\in\{1,k\}$, the peak signal strength which the receivers can handle can be different. This kind of problem, is not handled by traditional $L_1$ projection. Assuming that the lower bound is zero (e.g. for electrical signals), transmitting upper bounds can cost just 1 bit. The effective sparsity, can be improved by sending 1 bit for all the elements fixed at their corresponding bounds, and sending 8 bit real numbers for all the remaining non-zero entities as compared to sending 8 bit reals for all non-zero entities. 
We conjecture that the degree of freedom for the bounded \Lo problem is bounded below that for the unbounded \Lo problem, and hence can provide increased sparsity in terms of the bounds.

\begin{figure}[ht!]
\centering
  \includegraphics[width=8cm,height=4cm]{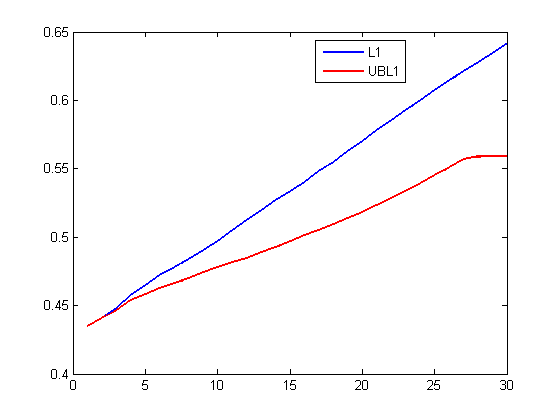}
  \caption{SURE criteria for \Lo (blue) and upper bounded \Lo (red). \label{Fig:df}
}
\end{figure}
Let us again assume the simple estimation problem $y = \Phi x + \epsilon$. Now for estimating the SURE criterion, we estimate $x$ and then generate the a new set of observation $\hat{y}$. The covariance between the terms $y$ and $\hat{y}$ is a scaled measure for the DF. Fig.~\ref{Fig:df} shows the covariance estimate for \Lo compared to upper bounded $L_1$. Upper bounded \Lo is uniformly bounded below the DF for only $L_1$. This also emphasizes the conjecture for bounded constrained $L_1$, the predicates which are fixed at their consecutive upper bound can be considered to fixed and as such do not contribute to the model complexity.

\section{Separable Quadratic Problems}
The problem of separable quadratic programming with linear bound constraints was first considered by Megiddo \etal~\cite{Megiddo93}. They proposed linear time solution to the generic problem by Lagrangian relaxation based on the multidimensional search procedure of~\cite{Megiddo84}. We introduce a novel linear time algorithm for gradient projection based norm minimization problem. Our starting point is an efficient method for projection onto the probabilistic simplex, with additional upper bound constraints. For infinite upper bound, this is the same method as proposed by numerous authors, namely Gafni et al.\cite{Gafni84}, Bertsekas~\cite{Bertsekas99}, Crammer et al.~\cite{Crammer00}, and more recently by ShalevSchwartz et al.~\cite{ShalevShwartz06} and Duchi et al.~\cite{Duchi08}. The basic intuition is that once the vector to be projected is ordered then the projection can be calculated exactly in linear time. Duchi et al.~\cite{Duchi08} proved the similarity of \Lo projection to the simplex projection, although the problem can be traced back to a special case of separable quadratic problem tackled by Megiddo \etal~\cite{Megiddo93}. Although the projection step is linear, the sorting/ordering step is still $O(n \log n)$. We contest that once the upper bounds are introduced, the ordering does not remain as simple as the previous works. In this paper we propose a linear time algorithm which orders the difference of the bounds from the gradient vector to compute the projections on the norm constraint.

\section{Upper bounded Simplex Projection}\label{SEC:UBSP}
The most basic projection task we consider can be formally described as the following optimization problem, for $\mathbf{v} \in \mathbb{R}^{n}_+$
\begin{equation}\label{Eqn:Min_generic}
    \min_{\mathbf{{x}}} \frac{1}{2}\|\mathbf{{x}-v}\|_2^2
  ~~\textrm{s.t.} ~~ \{\mathbf{x}\in \mathbs{\Omega}: \sum_{j=1}^n x_j \le z, ~\mathbf{0} \preccurlyeq \mathbf{x} \preccurlyeq \mathbf{b} \}
\end{equation}
where $\mathbf{0}\in {\mathbb{R}}^n$ is the vector of all zeros, $\mathbf{b}\in {\mathbb{R}}^n_+$ is the vector of upper bounds and $(\mathbf{p}\preccurlyeq \mathbf{q})$ denotes that $p_i \le q_i, ~\forall i\in\{1,n\}$.
Note that we enforce $\sum_{j=1}^n v_j \ge z$, because otherwise $\mathbf{x = v}$ is the optimal solution.
We assume that the feasibility of the constraints with respect to each other, and the existence of a solution is guaranteed by the set $\mathbs{\Omega}$ being non-empty.
Also note that if $z\mathbf{1}\preccurlyeq \mathbf{b}$ or $\mathbf{v}\preccurlyeq \mathbf{b}$, where $\mathbf{1}$ is the vector of all ones of size $n$, then the problem reduces to the projection onto the simplex problem of~\cite{Duchi08}.
\begin{claim}\label{Claim:x_is less_than_v}
$x_i \le v_i,~~\forall i\in\{1,2,\ldots,n\}$.
\end{claim}

The norm of $\mathbf{v}$ is larger than the constraint $z$ as mentioned earlier. Assume there is an optimal projection $\mathbf{x}^{\star}$, with its element $x^{\star}_i > v_i$. There exists another solution $\mathbf{\hat{x}}^{\star}$, such that $\hat{x}^{\star}_i = v_i$ and all other elements same as $\mathbf{x}^{\star}$, which is bounded and gives a lower value for the cost function, and hence is a better solution than $\mathbf{x}^{\star}$, which is a contradiction.

\begin{claim}\label{Claim:effective_lb}
$b_i^{effective} = \min (b_i, v_i),~~\forall i\in\{1,2,\ldots,n\}$.
\end{claim}

The maximum value that $x_i$ can reach is either the upper bound $b_i$ or $v_i$ (from Claim~\ref{Claim:x_is less_than_v}), hence the effective upper bound $b_i^{effective}$ is the minimum of $b_i$ or $v_i$.

From this point onwards, we will assume ${b_i} = {b_i}^{effective}~\forall i$, if not mentioned otherwise.
The next lemma is the extension of Lemma 1 of Duchi et al.~\cite{Duchi08} for bounded projections.
\begin{lemma}\label{Lemma:v_based_zero_constraint_ordering}
If $v_i > v_j$ and in optimal solution $x_i = 0$ then $x_j=0$, irrespective of ordering of $b_i$, $b_j$.
\end{lemma}
\emph{Proof.} We need to prove the above lemma for 2 cases.

[1] $b_i \ge b_j$. In this case, if another solution is constructed such that $x_i$ and $x_j$ are switched keeping all other indices same, then there is a strict decrease in optimal value, generating a contradiction.

[2] $b_i < b_j$. Again let us assume $x_j > 0$. Let us construct another optimal solution $\mathbf{\hat{x}}$, such that $\hat{x}_i = \Delta$, $\hat{x}_j = x_j - \Delta$, where $\Delta = \min(b_i, x_j)$, and keep all the other indices same. It can be easily observed that the norm as well as the upper bound constraint are satisfied for $\mathbf{\hat{x}}$. Now we can show that
\begin{align} \nonumber
 & \textrm{New obj value} - \textrm{Old obj value} \\ \nonumber
 &~=~ (v_i - \hat{x}_i)^2 + (v_j - \hat{x}_j)^2 - (v_i-x_i)^2 - (v_j-x_j)^2 \\ \nonumber
 &~=~ (v_i - \Delta )^2 + (v_j - x_j + \Delta )^2  - v_i^2 - (v_j-x_j)^2 \\ \nonumber
 &~=~ \Delta ^2 - 2v_i \Delta + \Delta ^2 + 2v_j \Delta - 2x_j \Delta \\ \nonumber
 &~=~ 2 \Delta ( \Delta - x_j ) + 2 ( v_j - v_i ) \Delta ~\le~ 2 ( v_j - v_i ) \Delta ~ < ~ 0
\end{align}
which is a contradiction since we constructed a solution better than the optimal solution.$~\square$

\section{Euclidean Projection onto the box-constrained \Lo Ball}\label{SEC:EPL1}
We modify the problem studied by Duchi et al.~\cite{Duchi08} to the more generic scenario containing the bounds on the predicted vector. We need to find the projection of a vector $\mathbf{v} \in \mathbb{R}^n$ onto a feasible region defined by
\begin{equation}\label{Eqn:Min}
      \min_{\mathbf{x}} \frac{1}{2}\|\mathbf{x-v}\|_2^2
  ~~\textrm{s.t.}~~ \{\mathbf{x}\in \mathbs{\Omega}: \|\mathbf{x}\|_1 \le z, ~\mathbf{a} \preccurlyeq \mathbf{x} \preccurlyeq \mathbf{b} \}
\end{equation}
Note that the vector $\mathbf{v}$ is no longer contained within the positive real space. We assume that $\mathbs{\Omega} \ne \varnothing$ guarantees feasibility and existence of a solution.
The range in which the $x_j$'s should lie can now be characterized as (a) $[a_j, b_j] < 0$, (b) $0 < [a_j, b_j]$ and (c) $a_j \le 0 \le b_j$, assuming that $a_j < b_j$ for all cases.

\textbf{Intervals not containing 0: }
Further analysis of the cost function in Eq.\eqref{Eqn:Min}, leads to the observation that conditions (a) and (b), are equivalent under a sign flip. This can be obtained by observing the following identities: (a) distance preservation under sign flip  $\|\mathbf{v-x}\|_2^2 = \|\mathbf{(-v)-(-x)}\|_2^2$, (b) 1 norm preservation under sign flip $\|\mathbf{x}\|_1 = \|\mathbf{-x}\|_1$ and (c) range transformation under sign flip$\mathbf{x}\in [\mathbf{a,b}]<0 \Leftrightarrow \mathbf{-x}\in [\mathbf{-b,-a}]>0$.
For such constraints, we can transform the bounds such that all the boundaries are positive.
Once this is done a simple change of variables
\begin{eqnarray}
  && \mathbf{\{v,x,b\} - a} = \mathbf{\hat{v},\hat{x},\hat{b}},~~ z - \|\mathbf{a}\|_1 = \hat{z}
\end{eqnarray}
leads to the formulation in Eq.\eqref{Eqn:Min} with the lower bound terms $a_i$'s equal to 0. 
The equivalent simpler problem is exactly similar to Eq.\eqref{Eqn:Min_generic}, since the $L_1$ constraint and the simplex constraint are same for positive variables. Also note that
element wise manipulations can be performed when all the bounds do \textit{not} belong to one particular case, without altering the form of the equations.

{\textbf{Interval containing 0: }}
The objective function in Eq.\eqref{Eqn:Min} has expression of the form $ \| \mathbf{v} - \mathbf{x} \| _p$ ($p=2$) and norm constraint is equivalent to inclusion in \Lp norm ball ($p=1$). Given a candidate solution $ \mathbf{x} \in \mathbb{R}^n $, let us define a category of moves which can be used to generate a family of candidate solutions.
By setting some subset of vector components to zero, we can generate corresponding points in all orthants, resulting in a family of  up to $2^n$ candidate solutions, one in each orthant. We note two properties of orthant projection move which are essential for our exposition. First, orthant projection move preserves \Lp norm ball inclusion constraint for all $p$. Second, under orthant projection moves of $ \mathbf{x} $, $ \| \mathbf{v} - \mathbf{x} \|_p$ is minimized when $ \mathbf{x} $ and $ \mathbf{v} $ are in the same orthant.  Together, these two properties ensure that there is a preferred orthant (determined apriori) where optimal solution is guaranteed to lie, as long as none of the other constraints are violated. This observation can be used to generalize Lemma 3 of Duchi et al.~\cite{Duchi08} to a much wider class of problems. For completion and ease of understanding we state the following lemma:
\begin{lemma}\label{Lemma:Duchi3}
\texttt{[Lemma 3, Duchi et al.\cite{Duchi08}]} Let $\mathbf{x}$ be an optimal solution of Eq.\eqref{Eqn:Min}. Then,
$x_i v_i \ge 0,~\forall i$.
\end{lemma}
Hence, $x_i$ has the same sign as $v_i$. Note that this lemma holds true for the upper bounded problem as well, since replacing any variable with 0 does not violate this constraint, and hence the proof for the lemma can be exactly applied to the more generic case mentioned above.

Orthant projection also reveals a possible failure mode for the above generalization. If there are terms other than those of the form of $ \| \mathbf{v} - \mathbf{x} \|_p$ and there are constraints other than those of the form of inclusion in \Lp norm ball, then apriori preferred orthant selection may not be possible.
Preferred orthant selection allows us to simplify the functional form of \Lp norm terms.
In particular, for \Lo norm projection problem, once problem has been transformed to guarantee that optimal solution lies in first orthant, \Lo norm constraint becomes equivalent to the simplex constraints, $x_i \ge 0 ~\forall i$ and $ \sum_{i=1}^n{x_i} \le z $.

In order to retain this simplification, any generalization involving terms which are not conducive to orthant projection must be explicitly handled. Box constraints are in general not conducive, unless interval contains origin.
For intervals such as $a_j < 0 < b_j$, we have the following claim:
\begin{claim}\label{Claim:a_b_v}
$|x_i| \in \begin{cases}
   \{0,|a|\} & v_i < 0 \\
   \{0,|b|\} & v_i > 0
  \end{cases}$
\end{claim}


Based on the above discussions we can write the generic upper bounded \Lo norm projection problem as, given any $\mathbf{\hat{v}} \in \mathbb{R}^n$, take the absolute value of the elements and transform it to $\mathbf{{v}} \in \mathbb{R}^n_+$, transform $[\mathbf{a,b}]\Rightarrow [\mathbf{0,{b}}]$ based on claim.~\ref{Claim:a_b_v}, find $\mathbf{{x}}$ by solving
\begin{equation}\label{Eqn:Min0}
      \min_{\mathbf{{x}}} \frac{1}{2}\|\mathbf{{x}-{v}}\|_2^2
 ~~\textrm{ s.t.}~~  \{\mathbf{{x}}\in \mathbs{\Omega}: \|\mathbf{{x}}\|_1 \le z, ~\mathbf{0} \preccurlyeq \mathbf{{x}} \preccurlyeq \mathbf{{b}} \}
\end{equation}
and return the final projection as $\mathbf{\hat{x}} = \mathbf{{x}}.*\textrm{sign}(\mathbf{\hat{v}})$.

Identifying the simplex projection as well as $L_1$ projection as the same problem leads us to study the unified problem mentioned above in Eq.\eqref{Eqn:Min0}.

The Lagrangian for the above optimization problem (Eq.\eqref{Eqn:Min0}) can be written as
\begin{equation}\label{Eqn:Lagrangian}
    \mathcal{L}= \frac{1}{2}\|\mathbf{x-v}\|_2^2 + \theta\left( \sum_{i=1}^n x_i - z \right) - \mathbs{\zeta}.\mathbf{x} - \mathbs{\gamma}.(\mathbf{b-x})
\end{equation}
Differentiating with respect to $x_i$ and comparing to zero gives the first order
optimality condition,
\begin{equation}\label{Eqn:first_order_opt}
    \frac{\partial \mathcal{L}}{\partial x_i} = x_i - v_i + \theta - \zeta_i + \gamma_i = 0
\end{equation}
The first complementary slackness KKT condition~\cite{Boyd04}, implies that
$x = 0$, when $v_i + \zeta_i = \theta_i$. Since, $\zeta_i > 0$, hence $x_i = 0$ whenever $v_i < \theta$.
The second complementary slackness KKT condition implies that
$0 < x_i < b_i$, means $\zeta_i=0$, $\gamma_i = 0$ and
\begin{equation}\label{Eqn:v_theta_gamma}
    x_i - v_i + \theta = 0
\end{equation}
The addition of finite upper bound leads to the next complementary slackness condition, namely, when the value of $x_i$ reaches it maximum value $b_i$, $\gamma_i > 0$ and
\begin{equation}\label{Eqn:v_theta_gamma_zero}
    b_i - v_i + \theta + \gamma_i = 0
\end{equation}

\begin{claim}\label{Claim:x_b_v}
$x_i = b_i$ implies $v_i \ge b_i$.
\end{claim}


Note that the converse is not generally true, that is $v_i > b_i$ does not imply that $x_i = b_i$, since it can still be lower than the upper bound.

\textbf{Corollary 4.}\label{Cor:x_eq_b} $v_i > b_i$ \textit{and} $\gamma_i > 0$ implies $x_i = b_i$.

One important aspect of the cost $\|\mathbf{x-v}\|_2^2$ is that the contribution of each $x_i$ to the total cost is dependent on the distance $v_i-x_i$. From this point onwards we will assume that the upper bound term $b_i = \min(v_i, b_i)$, such that $v_i - b_i \ge 0$. Since each $x_i$ is bounded to be less than $b_i$, hence $v_i-b_i$ can be thought to be the relative weight determining the order in which $x_i$'s should be changed to meet the norm constraint. This ordering can also be argued from the fact that the magnitude of the gradient with respect to $x_i$ is determined by the quantity $v_i-b_i$.

In Lemma~\ref{Lemma:v_based_zero_constraint_ordering}, we have shown that even for upper bounded simplex projection problem (after restriction to first orthant), such constraint ordering is possible for constraints $x_i \ge 0$, and is determined by $\mathbf{v}$.
In the next lemma 
we show that similar constraint ordering is possible for upper bound constraints $x_i \le b_i$, and is determined by $\mathbf{(v-b)}$ which is one of our key contributions and forms the basis of the proposed efficient algorithm. Based on the above observations we write a modified version of the lemma 2. from Shalev-Shwartz \etal~\cite{ShalevShwartz06}.
\begin{lemma}\label{Lemma:ub_constraint_order}
Let $\mathbf{x}$ be an optimal solution of Eq.\eqref{Eqn:Min0}. Let $i$ and $j$ be two indices such that $(v_i - b_i) \le (v_j - b_j)$. If $x_i = b_i$ then $x_j = b_j$ as well.
\end{lemma}
\emph{Proof.} From Eq.\eqref{Eqn:v_theta_gamma_zero}, whenever $x_i = b_i$, then $v_i-b_i = \theta + \gamma_i$ where $\gamma_i > 0$. Hence
\begin{align*}
              v_i-b_i &> \theta,~~~~~~~~~~~~~~~~~~~~~~~~\textrm{since} ~\gamma_i > 0 \\
  \Rightarrow v_j-b_j &\ge  v_i-b_i > \theta,~~~~~~~\textrm{given} \\
  \Rightarrow v_j-b_j &>  \theta \\
  \Rightarrow v_j-b_j &=  \theta + \gamma_j, ~~~~~~~~~~~~~~~\textrm{such that} ~\gamma_j > 0\\
  \Rightarrow x_j &= b_j~~~~~~~~~~~~~~~~~~~~~~~~\textrm{from Corollary 4}.~\square
\end{align*}

\subsection{Worst case strongly linear time algorithm}
We now propose an algorithm with strongly linear time worst case complexity which is asymptotically fastest possible. It is based on dependence between $\theta$ and $z$ along regularization path. We have already shown that, in optimal solution $x_i = 0$ whenever $v_i < \theta$, and $x_i = b_i$ whenever $v_i - b_i \ge \theta$ and equal to $v_i-\theta$ otherwise. 
Hence, for any value of $\theta$, variables $x_i$'s can be divided into three disjoint groups.
\begin{align}\label{Eqn:theta_to_xa}
 x_i =
  \begin{cases}
   0                 &\text{if}~v_i           \le \theta    ~~~~~~~~~~~~~~~~~~~\text{Fixed at lower limit} \\
   b_i              &\text{if}~v_i - b_i \ge \theta   ~~~~~~~~~~~\text{Fixed at upper limit}\\
   v_i - \theta &\text{if}~v_i - b_i < \theta < v_i  ~~\text{Constraints inactive} \\
  \end{cases}
\end{align}
Let us denote the sets of indices of $x_i$'s in these groups by $L$, $U$ and $C$ respectively. These sets are functions of $\theta$. Let optimal $\theta$ be $\theta^*$ and corresponding sets be $L^*$ $U^*$ and $C^*$. Relation between $z$ and $\theta$ can be expressed as,
\begin{equation}
z ~=~ \sum_{i=1}^n x_i =~ \sum_{i \in U} b_i + \sum_{i \in C}{ (v_i - \theta)} ~=~ \sum_{i \in U} b_i + \sum_{i \in C}{v_i} - | C | \theta \label{eqn:z_theta_relation_suboptimal}
\end{equation}
where $|.|$ for a set argument, denotes its cardinality. It is evident that $z$ is monotonically decreasing piece-wise linear function of $\theta$, with $2n$ points of discontinuity at $v_i$ and $(v_i - b_i)$ values.
The pseudo code of our proposed algorithm is given in Algorithm~\ref{Fig:Algo3}.
The algorithm operates upon merged $\mathbf{v}$ and $\mathbf{(v-b)}$ arrays, maintaining source information.

In the first stage, we find the linear segment corresponding to given $z$. Uncertainty interval $[\theta_L, \theta_R]$ for $\theta$ is initialized with  $[\min(\mathbf{v-b}),\max(\mathbf{v})]$ and is subsequently reduced in every iteration by bisection at a pivot selected from the elements of merged $\mathbf{v}$ and $\mathbf{(v-b)}$ arrays lying in current uncertainty interval.  For pivot, we use median, found using worst case linear time median finding algorithm~\cite{CLR01}, in order to ensure that number of iterations remains O($\log n$) and that after every iteration, size of uncertainty interval reduces by a constant fraction. Using Eq.\eqref{eqn:z_theta_relation_suboptimal} to evaluate $z$ at pivot $\theta$ is not efficient enough for overall linear time complexity, since summations involve O($n$) terms every time, resulting in O($n \log n)$ complexity. To rectify this inefficiency, apart from $S_{all} = \sum_{i=1}^n v_i$, we maintain two running partial sums across all iterations.

1) $S_L = $ sum of $v_i$ for all elements which are guaranteed to be set to zero in optimal solution i.e. $v_i \le \theta_L \Rightarrow i \in L^*$.

2) $S_R = $ sum of $(v_i - b_i)$ for all elements which are guaranteed to be set to corresponding upper bounds in optimal solution i.e. $(v_i - b_i) \ge \theta_R \Rightarrow i \in U^*$.

We also maintain cardinality  $n_L$ and $n_R$ of these sets. In terms of these partial sums, (\ref{eqn:z_theta_relation_suboptimal}) can be expressed as
\begin{eqnarray}\label{eqn:z_theta_relation_optimal}
    z_{pivot} &=& S_{all} - S_L - S_R  - ( n - n_L - n_R ) * \theta_{pivot} \\ \nonumber
    &-& \sum_{i : \theta_{pivot} \le v_i - b_i < \theta_R} ( v_i - b_i - \theta_{pivot} ) \\ \nonumber
    &-& \sum_{i : \theta_L < v_i \le \theta_{pivot}} ( v_i - \theta_{pivot} )
\end{eqnarray}
If $z_{pivot} > z_{target}$,  $[\theta_{pivot}, \theta_R]$ becomes new uncertainty interval, and $S_L$ and $n_L$ are updated as
\begin{equation}\label{eqn:S_L_update}
    \{S_L,n_L\} \leftarrow \{S_L,n_L\} +   \sum_{i : \theta_L < v_i \le \theta_{pivot}} \{v_i,1\}
\end{equation}
Otherwise, if $z_{pivot} < z_{target}$, $[\theta_L, \theta_{pivot}]$ becomes new uncertainty interval, and $S_R$, and $n_R$ are updated as
\begin{equation}\label{eqn:S_R_update}
    \{S_R,n_R\} \leftarrow \{S_R,n_R\} +   \sum_{i : \theta_{pivot} \le v_i - b_i < \theta_R} \{(v_i-b_i),1\}
\end{equation}
Iterations continue until there are no more points of discontinuity in the uncertainty interval and $L^*$, $U^*$ and $C^*$ have been found. Now, following modified version of (\ref{eqn:z_theta_relation_suboptimal}) can be used to evaluate $\theta^*$ as
\begin{equation}\label{eqn:theta_star}
\theta^* = \frac{\sum_{i \in U^*} b_i + \sum_{i \in C^*} v_i - z_{target}}{| C^* |}
\end{equation}

\begin{algorithm}
\caption{Algorithm for worst case strongly linear time projection onto the simplex with finite upper bound.}
\label{Fig:Algo3}
\begin{algorithmic}
\STATE REQUIRE $\mathbf{v}\in {\mathbb{R}}^n$,  $\mathbf{b} \in {\mathbb{R}}^n$, $ 0 < z_{target} < \sum_{i=1}^n b_i$
\STATE $\mathbf{vvb} \leftarrow merge(\mathbf{v,(v-b)})$ \COMMENT{maintain source info}
\STATE $(idx\_\theta_L, idx\_\theta_R) \leftarrow (0,2n-1)$
\STATE $S_{all} \leftarrow sum(\mathbf{v})$
\STATE $(n_L, S_L) \leftarrow 0 $
\STATE $(n_R, S_R) \leftarrow 0 $

\WHILE{ $idx\_\theta_R > idx\_\theta_L + 1$}

\STATE $\theta_{pivot}  \leftarrow pivot\_select( \mathbf{vvb}, idx\_\theta_L, idx\_\theta_R)$
\STATE $partition(\mathbf{vvb},idx\_\theta_L,idx\_\theta_R, \theta_{pivot})$
\STATE $idx\_\theta_{pivot} \leftarrow index(\theta_{pivot}) $

\STATE Evaluate $z_{pivot}$ using (\ref{eqn:z_theta_relation_optimal})

\IF{ $z_{pivot} > z_{target}$}
\STATE $idx\_\theta_L \leftarrow idx\_\theta_{pivot}$
\STATE Update $(S_L,n_L)$ using (\ref{eqn:S_L_update})
\ELSE
\STATE $idx\_\theta_R \leftarrow idx\_\theta_{pivot}$
\STATE Update $(S_R,n_R)$ using (\ref{eqn:S_R_update})
\ENDIF
\ENDWHILE
\STATE Evaluate $\theta^*$ using (\ref{eqn:theta_star})
\STATE RETURN $\mathbf{x}$ corresponding to $\theta^*$~(Eq.\ref{Eqn:theta_to_xa}).
\end{algorithmic}
\end{algorithm}

\section{Experiments}\label{SEC:Expts}
\begin{figure*}[htbp!]
\centering
  \includegraphics[width=8cm,height=5cm]{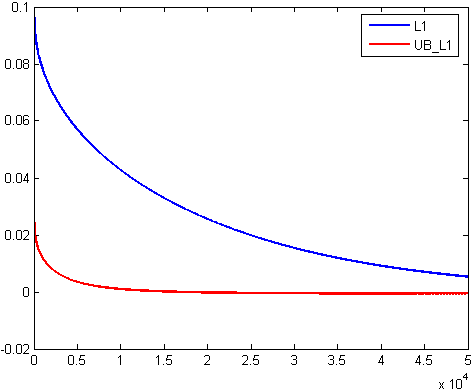}
  \includegraphics[width=8cm,height=5cm]{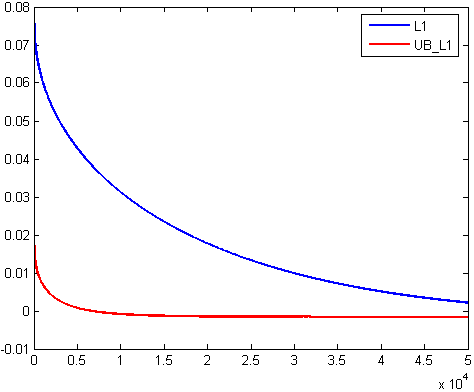}
  \includegraphics[width=16cm,height=5cm]{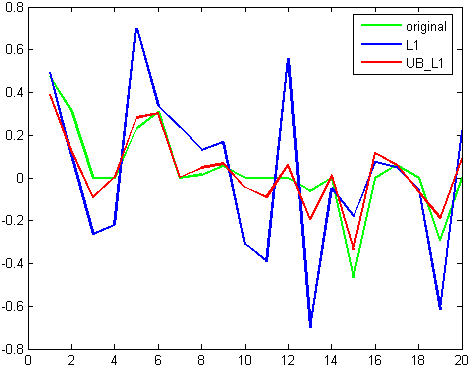}
  \caption{Top left: plot for abs$(f(\mathbf{w}) - f(\mathbf{w}^\star))$ vs. iteration (training). Top right: similar plot for testing. Bottom: estimated parameter vector $\mathbf{w}$. For all panels, blue: L1, red: our method UB\_L1 and green: ground truth. \label{Fig:comp}
}
\end{figure*}

The first set of experiments are performed for synthetic data. We generate labeled data belonging to 2 classes such that the probability of the label being 1/0 is distributed according to logistic likelihood 
$p(y_i=1|\mathbf{x_i,w})=\sigma(\mathbf{w.x_i})$,
where $\sigma(a) = 1/(1+\exp(-a))$. Additionally, we disturb 10\% of the data labels by introducing false labels based on random draws.
The ground truth parameter vector $\mathbf{w}$ is generated from a generalized Gaussian distribution, with rejection, such that the individual elements of the vector are bounded within $\pm 0.5$. Moreover half the entities of $w$ are made zeros to generate a sparse vector. The inference problem is thus an estimation problem with known upper bounds. 
We minimize the average logistic log loss, and project the gradient vector to the convex space. The norm constraint is determined as a fixed fraction of the dimension of the vector. The estimation error against the iterations, where the error is denoted as $f(\mathbf{w}) - f(\mathbf{w}^\star)$, for L1 and our method called UB\_L1 is shown in Fig.~\ref{Fig:comp}(left). 
Note that the L1 method estimates are outside the bounds which manifests itself as slower rate of convergence as evident from the plots. At convergence UB\_L1 estimate seems much closer to the ground truth parameter vector than the L1 estimate.

\begin{figure*}[htbp!]
\centering
  \includegraphics[width=16cm,height=5cm]{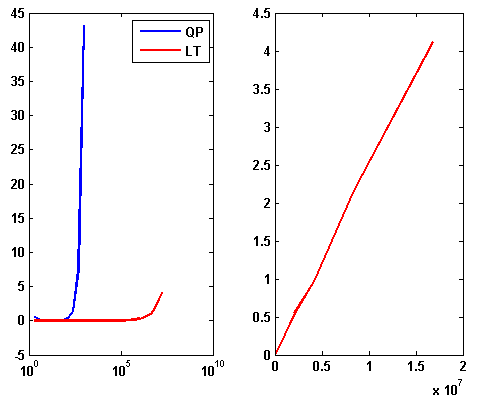}
  \caption{Left: comparison of our method against Matlab QP. Blue: run time (in seconds) for QP implementation of MATLAB. Red: run time for our linear time method. The horizontal axis runs over the dimension of the input vector $v$. Right: zoomed in red curve. \label{Fig:time_qp_algo2}
}
\end{figure*}
Next we explore the run time performance of our algorithm, against a standard quadratic programming (QP) implementation in MATLAB. Fig.~\ref{Fig:time_qp_algo2} shows the results for such an experiment. For projecting dense vectors with 1M non-zero
elements our method takes around 0.22 seconds.

\textbf{Food distribution}
The next experiment is drawn from a real world scenario and motivates the upper bound constraints.
We start by noting that the problem of food distribution can be easily applied to our case. Suppose the \textit{production} of one food item (e.g. chicken) in 40 states in the US is provided as the initial vector $\mathbf{v}$\footnote{http://www.agcensus.usda.gov/Publications/2007}. Assume that $r\%$ of the total production is put up for sales.

\begin{figure*}[htbp!]
\centering
  \includegraphics[width=14cm,height=6cm]{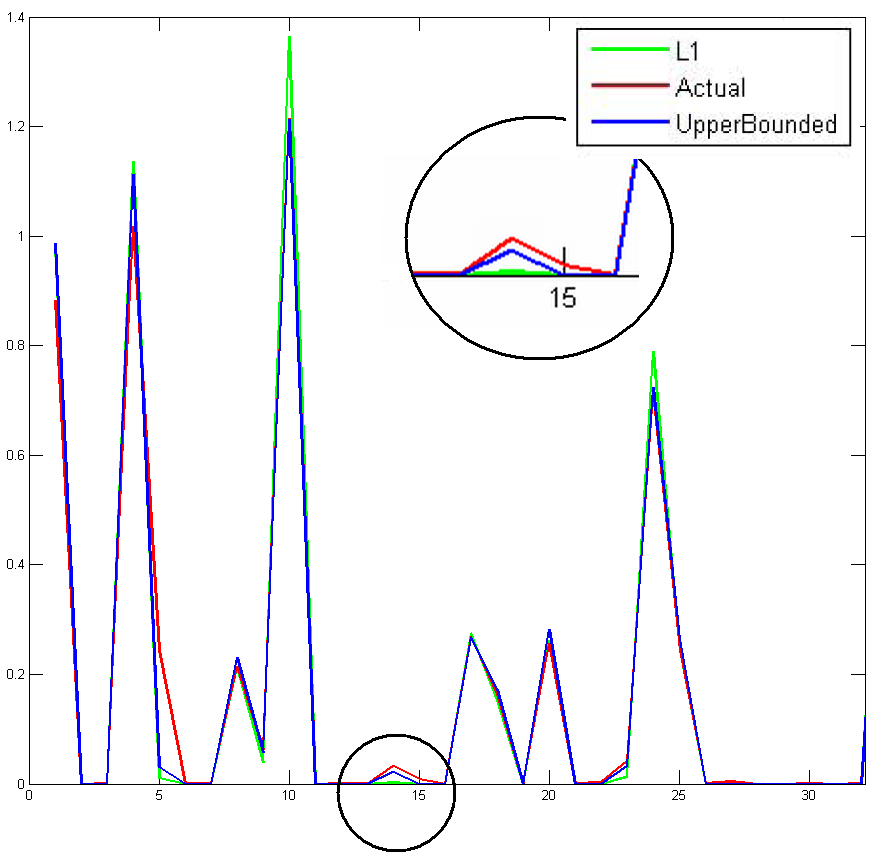}
  \caption{Red: Actual sales of Chicken in 40 states in 2007. Blue: our method with upper bounds. Green: $L1$ only. Note the small region in the circle which has been enlarged. $L1$ completely misses this region whereas our method still provides some value to it.\label{Fig:L1_comp_us}
}
\end{figure*}

We would like to find the \textit{sales} vector $\mathbf{x}$ for the 40 states. The upper bounds can be obtained from the consumption patterns in the previous years. We take production in 2007 as the new vector $\mathbf{v}$, and the distribution in 2004 as the upper bound $\mathbf{b}$. To remove scale differences the upper bound is normalized such that $\|\mathbf{b}\|_2=\|\mathbf{v}\|_2$. The norm constraint $z=\|\mathbf{v}\|_2*r/100$. The results for such an experiment are shown in Fig.~\ref{Fig:L1_comp_us}. As the value of $z$ decreases $L1$ forces more and more mass into the dominant elements. Our method still tries to satisfy the upper bound constraints, which spreads the distribution at the cost of sparsity. As the supply decreases, $L1$ tries to bias the distributions among the states based on the relative weights of the production itself. Our method, on the other hand, applies the demand based upper bounds, and biases the distribution in favor of the states with \textit{maximum disparity} between production and supply.

\section{Conclusion}\label{SEC:Concl}
In this paper we extend the idea of $L_1$ constrained gradient projection under the presence of upper bound constraints. We explore simplex projection with upper bounds and bring out the similarities with \Lo projection. We derive criteria for a-priori determination of sequence in which various constraints become active and use such orderings to propose an efficient algorithm. The key insight obtained from our experiments was that $L_1$ tries to increase the dominant elements while putting zeroes for all the others. Bound constrained $L_1$, weighs the elements based on their distance from the corresponding bound. This case leads to better predictions, specifically in cases which should be weighed based on the disparity between the demand and supply. The elements with higher disparity get higher weight in the predicted distribution vector.

{
\bibliographystyle{plain}
\bibliography{latex8}
}

\end{document}